\catcode`\@=11					



\font\fiverm=cmr5				
\font\fivemi=cmmi5				
\font\fivesy=cmsy5				
\font\fivebf=cmbx5				

\skewchar\fivemi='177
\skewchar\fivesy='60


\font\sixrm=cmr6				
\font\sixi=cmmi6				
\font\sixsy=cmsy6				
\font\sixbf=cmbx6				

\skewchar\sixi='177
\skewchar\sixsy='60


\font\sevenrm=cmr7				
\font\seveni=cmmi7				
\font\sevensy=cmsy7				
\font\sevenit=cmti7				
\font\sevenbf=cmbx7				

\skewchar\seveni='177
\skewchar\sevensy='60


\font\eightrm=cmr8				
\font\eighti=cmmi8				
\font\eightsy=cmsy8				
\font\eightit=cmti8				
\font\eightbf=cmbx8				

\skewchar\eighti='177
\skewchar\eightsy='60


\font\ninei=cmmi9
\font\ninesy=cmsy9

\skewchar\ninei='177
\skewchar\ninesy='60


\font\tenrm=cmr10				
\font\teni=cmmi10				
\font\tensy=cmsy10				
\font\tenex=cmex10				
\font\tenit=cmti10				
\font\tensl=cmsl10				
\font\tenbf=cmbx10				
\font\tentt=cmtt10				
\font\tenss=cmss10				
\font\tensc=cmcsc10				
\font\tenbi=cmmib10				

\skewchar\teni='177
\skewchar\tenbi='177
\skewchar\tensy='60

\def\tenpoint{\ifmmode\err@badsizechange\else
	\textfont0=\tenrm \scriptfont0=\sevenrm \scriptscriptfont0=\fiverm
	\textfont1=\teni  \scriptfont1=\seveni  \scriptscriptfont1=\fivemi
	\textfont2=\tensy \scriptfont2=\sevensy \scriptscriptfont2=\fivesy
	\textfont3=\tenex \scriptfont3=\tenex   \scriptscriptfont3=\tenex
	\textfont4=\tenit \scriptfont4=\sevenit \scriptscriptfont4=\sevenit
	\textfont5=\tensl
	\textfont6=\tenbf \scriptfont6=\sevenbf \scriptscriptfont6=\fivebf
	\textfont7=\tentt
	\textfont8=\tenbi \scriptfont8=\seveni  \scriptscriptfont8=\fivemi
	\def\rm{\tenrm\fam=0 }%
	\def\it{\tenit\fam=4 }%
	\def\sl{\tensl\fam=5 }%
	\def\bf{\tenbf\fam=6 }%
	\def\tt{\tentt\fam=7 }%
	\def\ss{\tenss}%
	\def\sc{\tensc}%
	\def\bmit{\fam=8 }%
	\rm\setparameters\setbaselines\fi}


\font\twelverm=cmr12				
\font\twelvei=cmmi12				
\font\twelvesy=cmsy10	scaled\magstep1		
\font\twelveex=cmex10	scaled\magstep1		
\font\twelveit=cmti12				
\font\twelvesl=cmsl12				
\font\twelvebf=cmbx12				
\font\twelvett=cmtt12				
\font\twelvess=cmss12				
\font\twelvesc=cmcsc10	scaled\magstep1		
\font\twelvebi=cmmib10	scaled\magstep1		

\skewchar\twelvei='177
\skewchar\twelvebi='177
\skewchar\twelvesy='60

\def\twelvepoint{\ifmmode\err@badsizechange\else
	\textfont0=\twelverm \scriptfont0=\eightrm \scriptscriptfont0=\sixrm
	\textfont1=\twelvei  \scriptfont1=\eighti  \scriptscriptfont1=\sixi
	\textfont2=\twelvesy \scriptfont2=\eightsy \scriptscriptfont2=\sixsy
	\textfont3=\twelveex \scriptfont3=\tenex   \scriptscriptfont3=\tenex
	\textfont4=\twelveit \scriptfont4=\eightit \scriptscriptfont4=\sevenit
	\textfont5=\twelvesl
	\textfont6=\twelvebf \scriptfont6=\eightbf \scriptscriptfont6=\sixbf
	\textfont7=\twelvett
	\textfont8=\twelvebi \scriptfont8=\eighti  \scriptscriptfont8=\sixi
	\def\rm{\twelverm\fam=0 }%
	\def\it{\twelveit\fam=4 }%
	\def\sl{\twelvesl\fam=5 }%
	\def\bf{\twelvebf\fam=6 }%
	\def\tt{\twelvett\fam=7 }%
	\def\ss{\twelvess}%
	\def\sc{\twelvesc}%
	\def\bmit{\fam=8 }%
	\rm\setparameters\setbaselines\fi}


\font\fourteenrm=cmr12	scaled\magstep1		
\font\fourteeni=cmmi12	scaled\magstep1		
\font\fourteensy=cmsy10	scaled\magstep2		
\font\fourteenex=cmex10	scaled\magstep2		
\font\fourteenit=cmti12	scaled\magstep1		
\font\fourteensl=cmsl12	scaled\magstep1		
\font\fourteenbf=cmbx12	scaled\magstep1		
\font\fourteentt=cmtt12	scaled\magstep1		
\font\fourteenss=cmss12	scaled\magstep1		
\font\fourteensc=cmcsc10 scaled\magstep2	
\font\fourteenbi=cmmib10 scaled\magstep2	

\skewchar\fourteeni='177
\skewchar\fourteenbi='177
\skewchar\fourteensy='60

\def\fourteenpoint{\ifmmode\err@badsizechange\else
	\textfont0=\fourteenrm \scriptfont0=\tenrm \scriptscriptfont0=\sevenrm
	\textfont1=\fourteeni  \scriptfont1=\teni  \scriptscriptfont1=\seveni
	\textfont2=\fourteensy \scriptfont2=\tensy \scriptscriptfont2=\sevensy
	\textfont3=\fourteenex \scriptfont3=\tenex \scriptscriptfont3=\tenex
	\textfont4=\fourteenit \scriptfont4=\tenit \scriptscriptfont4=\sevenit
	\textfont5=\fourteensl
	\textfont6=\fourteenbf \scriptfont6=\tenbf \scriptscriptfont6=\sevenbf
	\textfont7=\fourteentt
	\textfont8=\fourteenbi \scriptfont8=\tenbi \scriptscriptfont8=\seveni
	\def\rm{\fourteenrm\fam=0 }%
	\def\it{\fourteenit\fam=4 }%
	\def\sl{\fourteensl\fam=5 }%
	\def\bf{\fourteenbf\fam=6 }%
	\def\tt{\fourteentt\fam=7}%
	\def\ss{\fourteenss}%
	\def\sc{\fourteensc}%
	\def\bmit{\fam=8 }%
	\rm\setparameters\setbaselines\fi}


\font\seventeenrm=cmr10 scaled\magstep3		


\newdimen\rp@
\newcount\@basestretchnum
\newskip\@baseskip
\newskip\headskip
\newskip\footskip


\def\setparameters{\rp@=.1em
	\headskip=24\rp@
	\footskip=\headskip
	\delimitershortfall=5\rp@
	\nulldelimiterspace=1.2\rp@
	\scriptspace=0.5\rp@
	\abovedisplayskip=10\rp@ plus3\rp@ minus5\rp@
	\belowdisplayskip=10\rp@ plus3\rp@ minus5\rp@
	\abovedisplayshortskip=5\rp@ plus2\rp@ minus4\rp@
	\belowdisplayshortskip=10\rp@ plus3\rp@ minus5\rp@
	\normallineskip=\rp@
	\lineskip=\normallineskip
	\normallineskiplimit=0pt
	\lineskiplimit=\normallineskiplimit
	\jot=3\rp@
	\setbox0=\hbox{\the\textfont3 B}\p@renwd=\wd0
	\skip\footins=12\rp@ plus3\rp@ minus3\rp@
	\skip\topins=0pt plus0pt minus0pt}


\def\setbaselines{\maxdepth=4\rp@\baselinestretch=\@basestretchnum}


\def\baselinestretch{\afterassignment\@basestretch\@basestretchnum}
\def\@basestretch{%
	\@baseskip=12\rp@ \divide\@baseskip by1000
	\normalbaselineskip=\@basestretchnum\@baseskip
	\baselineskip=\normalbaselineskip
	\bigskipamount=\the\baselineskip
		plus.25\baselineskip minus.25\baselineskip
	\medskipamount=.5\baselineskip
		plus.125\baselineskip minus.125\baselineskip
	\smallskipamount=.25\baselineskip
		plus.0625\baselineskip minus.0625\baselineskip
	\setbox\strutbox=\hbox{\vrule height.708\baselineskip
		depth.292\baselineskip width0pt }}



\def\makeheadline{\vbox to0pt{\baselinestretch=1000
	\vskip-\headskip \vskip1.5pt
	\line{\vbox to\ht\strutbox{}\the\headline}\vss}\nointerlineskip}

\def\makefootline{\baselineskip=\footskip\line{\the\footline}}

\def\big#1{{\hbox{$\left#1\vbox to8.5\rp@ {}\right.\n@space$}}}
\def\Big#1{{\hbox{$\left#1\vbox to11.5\rp@ {}\right.\n@space$}}}
\def\bigg#1{{\hbox{$\left#1\vbox to14.5\rp@ {}\right.\n@space$}}}
\def\Bigg#1{{\hbox{$\left#1\vbox to17.5\rp@ {}\right.\n@space$}}}


\mathchardef\alpha="710B
\mathchardef\beta="710C
\mathchardef\gamma="710D
\mathchardef\delta="710E
\mathchardef\epsilon="710F
\mathchardef\zeta="7110
\mathchardef\eta="7111
\mathchardef\theta="7112
\mathchardef\iota="7113
\mathchardef\kappa="7114
\mathchardef\lambda="7115
\mathchardef\mu="7116
\mathchardef\nu="7117
\mathchardef\xi="7118
\mathchardef\pi="7119
\mathchardef\rho="711A
\mathchardef\sigma="711B
\mathchardef\tau="711C
\mathchardef\upsilon="711D
\mathchardef\phi="711E
\mathchardef\chi="711F
\mathchardef\psi="7120
\mathchardef\omega="7121
\mathchardef\varepsilon="7122
\mathchardef\vartheta="7123
\mathchardef\varpi="7124
\mathchardef\varrho="7125
\mathchardef\varsigma="7126
\mathchardef\varphi="7127
\mathchardef\imath="717B
\mathchardef\jmath="717C
\mathchardef\ell="7160
\mathchardef\wp="717D
\mathchardef\partial="7140
\mathchardef\flat="715B
\mathchardef\natural="715C
\mathchardef\sharp="715D


\def\err@badsizechange{%
	\immediate\write16{--> Size change not allowed in math mode, ignored}}

\baselinestretch=1000
\tenpoint

\catcode`\@=12					
\catcode`\@=11
\expandafter\ifx\csname @iasmacros\endcsname\relax
	\global\let\@iasmacros=\par
\else	\immediate\write16{}
	\immediate\write16{Warning:}
	\immediate\write16{You have tried to input iasmacros more than once.}
	\immediate\write16{}
	\endinput
\fi
\catcode`\@=12


\def\rmb{\seventeenrm}

\def\singlespace{\baselineskip=\normalbaselineskip}
\def\halfspace{\baselineskip=1.5\normalbaselineskip}
\def\doublespace{\baselineskip=2\normalbaselineskip}


\def\AB{\bigskip\parindent=40pt
        \centerline{\bf ABSTRACT}\medskip\halfspace\narrower}
\def\AE{\bigskip\nonarrower\doublespace}
\def\nonarrower{\advance\leftskip by-\parindent
	\advance\rightskip by-\parindent}


\def\boxit#1{\vbox{\hrule\hbox{\vrule\kern3pt
	\vbox{\kern3pt#1\kern3pt}\kern3pt\vrule}\hrule}}

\def\hence{\leavevmode\hbox{\bf .\raise5.5pt\hbox{.}.} }

\def\dalemb#1#2{{\vbox{\hrule height.#2pt
	\hbox{\vrule width.#2pt height#1pt \kern#1pt \vrule width.#2pt}
	\hrule height.#2pt}}}
\def\gtorder{\mathrel{\raise.3ex\hbox{$>$}\mkern-14mu
             \lower0.6ex\hbox{$\sim$}}}
\def\ltorder{\mathrel{\raise.3ex\hbox{$<$}\mkern-14mu
             \lower0.6ex\hbox{$\sim$}}}

\newdimen\fullhsize
\newbox\leftcolumn
\def\twoup{\hoffset=-.5in \voffset=-.25in
  \hsize=4.75in \fullhsize=10in \vsize=6.9in
  \def\fullline{\hbox to\fullhsize}
  \let\lr=L
  \output={\if L\lr
        \global\setbox\leftcolumn=\columnbox\global\let\lr=R \advancepageno
      \else \doubleformat \global\let\lr=L\fi
    \ifnum\outputpenalty>-20000 \else\dosupereject\fi}
  \def\doubleformat{\shipout\vbox{
    \fullline{\box\leftcolumn\hfil\columnbox}\advancepageno}}
  \def\columnbox{\leftline{\vbox{\makeheadline\pagebody\makefootline}}}
  \tolerance=1000 }
\twelvepoint
\doublespace
{{
\rightline{~~~December, 2004}
\bigskip\bigskip
\centerline{\rmb Remarks on the History of Quantum Chromodynamics}
}
\medskip
\centerline{\it  Stephen L. Adler
}
\centerline{\bf Institute for Advanced Study}
\centerline{\bf Princeton, NJ 08540}
\medskip
\bigskip\bigskip
\leftline{\it Send correspondence to:}
\medskip
{\singlespace\leftline{Stephen L. Adler}
\leftline{Institute for Advanced Study}
\leftline{Einstein Drive, Princeton, NJ 08540}
\leftline{Phone 609-734-8051; FAX 609-924-8399; email adler@ias.
edu}}
\bigskip\bigskip

\vfill\eject
\pageno=2
\AB
I make some remarks on events leading to the final formulation of
quantum chromodynamics, stimulated by the ``Search and Discovery'' article
by Bertram Schwarzschild in the December, 2004 {\it Physics Today}.
The following text with references is being submitted as a letter
to the editor of {\it Physics Today}.
\AE
\bigskip\bigskip
\vfill\eject
\pageno=3
I read with appreciation Bertram Schwarzschild's article about the
richly deserved Nobel Prize won by David Gross, David Politzer, and
Frank Wilczek for the discovery of asymptotic freedom.  I am writing to
point out  significant inaccuracies and omissions in the
historical account that Schwarzschild
gives of the developments leading up to this work .  Schwarzschild
skips over important stages in the development of quantum chromodynamics
by confusing scaling results obtained in 1969 with current algebra sum
rules obtained four years earlier.  Gell-Mann's current algebra
was a set of algebraic relations between currents, abstracted
from a constituent quark model
for hadrons, with the aim of allowing calculations of relations
among the electromagnetic
and weak processes coupling to these currents, without requiring
details of the then unknown dynamics of the quarks.
The principal sum rules testing aspects of the Gell-Mann current algebra
were derived in 1965.   The first, which depended only
on the integrated axial-vector charge commutator, together with the PCAC
(partially conserved axial current) hypothesis, was the ``Adler--Weisberger''
sum rule (and equivalent soft pion theorem)
derived independently by  Weisberger [1] and by me [2,3]
in 1965.  This
related the nucleon axial-vector beta decay coupling $g_A$ to pion nucleon
scattering cross sections, and was in good accord with experiment, giving
great encouragement to the current algebra program. Many people entered the
field, and a variety of experimentally verified current algebra/PCAC soft
pion theorems were found.
In my longer article about the $g_A$ sum rule [3], I noted that, by using my
earlier observation [4] that forward neutrino reactions couple only to the
divergences of the weak currents, the PCAC assumption could be eliminated.
This led to relations involving cross sections for neutrino scattering
with a forward-going lepton, that provided exact tests of the integrated
charge commutation algebra.  Soon afterwards, during a visit to CERN in
the summer of 1965, Gell-Mann asked me whether I could make some comparable
statement about the local current algebra.
After considerable hard algebra, I discovered a sum rule [5] involving
structure functions in deep inelastic neutrino scattering that directly
tested the local Gell-Mann algebra.  At zero momentum transfer
squared $q^2=0$, the axial-vector part of the neutrino
sum rule reduced to the neutrino scattering form [3] of
the Adler-Weisberger sum rule, and near
$q^2=0$, the vector part reduced to the sum rule obtained by
Cabibbo and Radicatti [6] and others.
My sum rule for neutrino scattering was soon afterwards converted into an
inequality for deep inelastic electron scattering structure functions
by Bjorken [7].

Although not directly tested
until many years later [8], the neutrino sum rule had important conceptual
implications that figured prominently in developments over the next few years.
To begin with, it gave the first indications that
deep inelastic lepton scattering could give information about the local
properties of currents, a fact that at first seemed astonishing, but which
turned out to have important extensions.  Secondly, as noted by Chew in
remarks at the 1967 Solvay Conference [9], the closure property tested in
the sum rules, if verified experimentally, would suggest the presence of
elementary constituents inside hadrons.  In a Letter [10] published shortly
after this conference, Chew argued that my sum rule, if verified, would
rule out the then popular ``bootstrap'' models of hadrons, in which all
strongly interacting particles were asserted to be equivalent (``nuclear
democracy'').  In his words, ``such sum rules may allow confrontation
between an underlying local spacetime structure for strong interactions
and a true bootstrap.  The pure bootstrap idea, we suggest,
may be incompatible with closure.''  In a similar vein, Bjorken,
in his 1967 Varenna lectures [11],
argued that the neutrino sum rule was strongly suggestive of the presence
of hadronic constituents.

These conceptual developments still left undetermined the
mechanism by which the neutrino sum rule, and Bjorken's electron scattering
inequality, could be saturated at large $q^2$.  In an analysis that I
carried out with Gilman in 1966 of the saturation of the neutrino sum
rule for small $q^2$ [12], we pointed out that saturation
of the neutrino sum rule for large four-momentum transfer $q^2$ would
require a new
component in the deep inelastic cross section, that did not fall off
with form-factor squared behavior.  Bjorken became interested in the
issue of how the sum rule could be saturated, and formulated several
preliminary models that (in retrospect) already had hints of the dominance
of a regime where the energy transfer $\nu$ grows proportionately
to the value of $q^2$.
I summarized these pre-scaling
proposals of Bjorken in the discussion period of the 1967 Solvay
Conference [13] (which Bjorken did not attend),
in response to  questions from Chew and others
as to how the neutrino sum rule could be saturated.  The precise saturation
mechanism was clarified some months later with the proposal by Bjorken [14]
of scaling, and soon afterwards, with
the experimental work at SLAC [15] on deep inelastic electron scattering.

The Bjorken scaling hypothesis, together with parton model ideas that
were inspired by Feynman, led to powerful tools for studying deep inelastic
scattering that greatly extended the scope of what could be obtained using
only the Gell-Mann current algebra, precisely
because more specific dynamical input was assumed.
After the advent of the  scaling hypothesis,
Callan and Gross [16] used it to derive a proportionality
relation between two of the deep inelastic structure functions, under the
assumption of dominance  by spin-1/2 constituents
(partons in the later terminology), which
was testable in electroproduction as well as neutrino experiments, without
resort to the evaluation of sum rules. The Callan--Gross relation was one of
a number of parton model relations that went beyond the results obtainable
from current algebra. Within the parton model framework, the older current
algebra results also received a new interpretation; for example,
my neutrino sum rule could be recast as an integral over
the partonic density of the third component of isospin, which is independent
of $q^2$ because the proton is in an isospin 1/2 eigenstate.

Shortly after the Callan-Gross paper appeared, Tung and I [17], and
independently Jackiw and Preparata [18], showed that in
perturbation theory for
quantum field theory there
would be logarithmic deviations from the Callan-Gross relation. In other
words, only free field theory would give exact scaling.  In the memorable
words of a seminar talk by
Gell-Mann [19], in which he discussed work on light cone current algebra
that he carried out with Fritzsch [20],
``Nature reads the books of free  field theory.''
Recognition of this, together with the new renormalization group methods
of Wilson, Callan, and Symanzik discussed in Schwarzschild's article,
set the stage for
a search for field theories that would have almost free behavior, with the
resulting discovery of asymptotic freedom of Yang-Mills theories as the only
case that worked.

I also want to comment on the origins of the color hypothesis, leading
 up to  its
modern form -- the tripling of the number of  fractionally charged quarks
-- which was proposed as the solution to the wave function symmetry problem
in the seminal paper of Bardeen, Gell-Mann, and
Fritzsch [21].   (It was this paper
that introduced  the term ``color'' and marshaled additional experimental
evidence,
from neutral pion decay, and the hadron to muon production
ratio at $e^+e^-$ colliders, in its support.)
In 1969 I
gave a talk at the International Conference on High Energy Physics and
Nuclear Structure, reviewing the consequences of the axial-vector
anomaly [22] for
$\pi^0 \to \gamma \gamma$ decay, and as one of my closing remarks, I noted
that whereas the fractionally charged quark model gave a decay amplitude
a factor of 3 too small, the Han-Nambu model [23] with
three triplets of integrally charged quarks
supplied the missing factor of
3, giving a result in accord with experiment.  At the end of my talk, a
Russian physicist in the audience (I don't remember who)
came up and told me that Tavkhelidzde
had also proposed tripling the quark degrees of freedom, and asked me to
include the reference in the published version of my talk [24], which I did.
Tavkhelidze's paper, a published conference talk [25],
dealt mainly with quark model
mass and magnetic moment relations, but also noted that the $S$-state
wavefunction problem could be solved
``if we introduce additional quantum numbers which antisymmetrize the
total wave function.  Employing these additional quantum numbers we are able
to make the quark charges integer without violating the relations between
the magnetic moments.''  A similar observation was made in a paper of
Miyamoto [26], cited in a later review by Tavkhelidze [27].

Of the papers with triplets of integrally charged quarks, only the
Han-Nambu paper contemplated a possible dynamical role for octet
vector gluons, and only this paper was widely known in the U. S.
For example, Gell-Mann, in his plenary talk at the 1972 Fermilab
(then National Accelerator Laboratory) conference [28], in which
he discussed fractionally charged colored quarks as a
simplification of Greenberg's [29] parastatistics proposal, also
mentions the alternative of integrally charged Han-Nambu quarks.
In this same conference proceedings, as noted in a recent
historical article of Wilson [30] (see also Fritzsch [31]), the
first proposal of what we now know as QCD was given in the
parallel session talk of Fritzsch and Gell-Mann [32]. This paper,
in discussing  the dynamical implications of color octet gluons,
states: ``If the gluons become a color octet, then we do not have
to deal with a gluon field strength standing alone, only with its
square, summed over the octet, and with quantities like $\overline
q(\partial_{\mu}- ig\rho_A B_{A\mu}) q$, where the $\rho$'s are
the eight 3x3 color matrices for the quark and the B's are the
eight gluon potentials.''  Although the word Lagrangian is not
mentioned, this is a complete description of the two terms that
make up the QCD Lagrangian.  Within a year after this talk,
through the work of Gross, Politzer, and Wilczek, QCD was
enthroned as the candidate field theory for the strong
interactions.

In conclusion, I wish to acknowledge conversations with Tian-Yu Cao,
and correspondence with
Michela Massimi, that prompted me to look back at historical events leading
up to the final formulation of quantum chromodynamics.
This work was supported in part by the Department of Energy under
Grant \#DE--FG02--90ER40542.

\bigskip
\noindent
{ REFERENCES}
\bigskip
\noindent
[1] W. I. Weisberger, Phys. Rev. Lett. {\bf 14}, 1047 (1965); Phys. Rev.
{\bf 143}, 1302 (1966).   \hfill \break
\bigskip
\noindent
[2] S. L. Adler, Phys. Rev. Lett. {\bf 14}, 1051 (1965). The various
derivations of the $g_A$ sum rule given in refs. [1] and [2] included methods
that used the infinite momentum frame limit
introduced by S. Fubini
and G. Furlan, {\it Physics} {\bf 1}, 229 (1965).  \hfill\break
\bigskip
\noindent
[3] S. L. Adler,  Phys. Rev. {\bf 140}, B736 (1965).  \hfill\break
\bigskip
\noindent
[4]  S. L. Adler, Phys. Rev.  {\bf 135}, B963 (1964).\hfill\break
\bigskip
\noindent
[5]  S. L. Adler, Phys. Rev. {\bf 143}, 1144 (1966). This paper presented
three sum rules without addressing the issue of convergence; the convergent
$\beta$ sum rule is the one discussed here. The other two sum rules (as I
suspected at the time) were shown to be divergent, and so did not provide
tests of the Gell-Mann algebra.
\hfill\break
\bigskip
\noindent
[6]  N. Cabibbo and L. Radicati, Phys. Rev. Lett. {\bf 19}, 697 (1966).
\hfill\break
\bigskip
\noindent
[7]  J. D. Bjorken, Phys. Rev. Lett. {\bf 16}, 408 (1966).\hfill\break
\bigskip
\noindent
[8]  D. Allasia et al, {\it Z. Phys. C} {\bf 28}, 321 (1985).\hfill\break
\bigskip
\noindent
[9]  {\it Fundamental Problems in Elementary Particle Physics} (Proceedings
of the Fourteenth Conference on Physics at the University of Brussels,
October, 1967), Interscience Publishers, London, 1968.  Chew's remark, in
response to my discussion of Bjorken's models for saturation of the neutrino
sum rule, is on page 212 of this proceedings. \hfill\break
\bigskip
\noindent
[10] G. F. Chew, Phys. Rev. Lett. {\bf 19}, 1492 (1967).  Chew also refers to
a more general local current algebra sum rule derived by Fubini, the
integrand of which is not expressible in terms of measurable structure
functions.  See S. Fubini, {\it Nuovo Cimento} {\bf 43} A, 475 (1966), based
on methods of
S. Fubini, G. Furlan, and C. Rossetti, {\it Nuovo Cimento} {\bf 40} A,
1171 (1965).
\hfill\break
\bigskip
\noindent
[11]  J. D. Bjorken, Current Algebra at Small Distances, in {\it Proceedings
of the International School of Physics ``Enrico Fermi'' Course XLI},
J. Steinberger, ed., Academic Press, New York, 1968. See page 56 of this
proceedings. \hfill\break
\bigskip
\noindent
[12] S. L. Adler and F. J. Gilman, Phys. Rev. {\bf 156},
1598 (1967).\hfill\break
\bigskip
\noindent
[13]  S. L. Adler, untitled remarks on experimental tests of local current
algebra, in the Solvay Conference proceedings of ref. [9] above, pp. 205-214.
In these remarks I attributed the saturation models to Bjorken, but there
is no preprint reference; I believe I learned of the models directly from
Bjorken when we were both lecturers at the Varenna summer school in July,
1967 -- see the remarks on page 63 of his lectures cited in
ref. [11] above.  \hfill\break
\bigskip
\noindent
[14]  J. D. Bjorken, Phys. Rev. {\bf 179},  1547 (1969).\hfill\break
\bigskip
\noindent
[15]  R. E. Taylor, Rev. Mod. Phys. {\bf 63}, 573 (1991); H. W. Kendall,
Rev. Mod. Phys. {\bf 63}, 597 (1991); J. I. Friedman, Rev. Mod. Phys.
{\bf 63}, 615 (1991).\hfill\break
\bigskip
\noindent
[16]  C. G. Callan, Jr. and D. J. Gross, Phys. Rev. Lett.
{\bf 22}, 156 (1969). \hfill\break
\bigskip
\noindent
[17]  S. L. Adler and W.-K. Tung, Phys. Rev. Lett. {\bf 22}, 978 (1969).
\hfill\break
\bigskip
\noindent
[18]  R. Jackiw and G. Preparata, Phys. Rev. Lett. {\bf 22}, 975 (1969).
\hfill\break
\bigskip
\noindent
[19]  M. Gell-Mann, talk given at the Institute for Advanced Study,
Princeton, 1971 (unpublished).  See reference 10 of D. J. Gross and
S. B. Treiman, Phys. Rev. D {\bf 4}, 1059 (1971).\hfill\break
\bigskip
\noindent
[20]  H. Fritzsch and M. Gell-Mann, Light Cone Current Algebra,
talk at the 1971 Coral Gables Conference,
extended into a preprint a few months later; reissued recently as
arXiv: hep-ph/0301127. \hfill\break
\bigskip
\noindent
[21]  W. A. Bardeen, H. Fritzsch, and M. Gell-Mann, Light-Cone Current
Algebra, $\pi^0$ Decay, and $e^+e^-$ Annihilation, in {\it Scale and
Conformal Symmetry in Hadron Physics}, R. Gatto, ed., Wiley, New York
(1973); reissued recently as arXiv:  hep-ph/0212183.\hfill\break
\bigskip
\noindent
[22]  S. L. Adler, Phys. Rev. {\bf 177}, 2426 (1969; J. S. Bell and
R. Jackiw, Nuovo Cimento {\bf 60} A, 47 (1969); S. L. Adler and W. A.
Bardeen, Phys. Rev. {\bf 182},  1517 (1969).\hfill\break
\bigskip
\noindent
[23]  M. Y. Han and Y. Nambu, Phys. Rev. {\bf 139},   B1006 (1965).
\hfill\break
\bigskip
\noindent
[24]  S. L. Adler, $\pi^0$ Decay, in {\it High-Energy Physics and
Nuclear Structure}, Proceedings of the Third International Conference
on High Energy Physics and Nuclear Structure, S. Devons, ed., Plenum Press,
New York (1970), pp. 647-655.  See the remarks on p. 654.
\hfill\break
\bigskip
\noindent
[25]  A. Tavkhelidze, Higher Symmetries and Composite Models of Elementary
Particles, in {\it High Energy Physics and Elementary Particles},
International Atomic Energy Agency, Vienna (1965), pp. 753-762.
The quoted remark is on p. 758. \hfill\break
\bigskip
\noindent
[26] Y. Miyamoto, Three Kinds of Triplet Model,
in {\it Extra Number Supplement of Progress of Theoretical
Physics: Thirtieth Anniversary of the Yukawa Meson Theory} (1965),
p. 187.  \hfill\break
\bigskip
\noindent
[27]  A. N. Tavkhelidze, Color, Colored Quarks, Quantum Chromodynamics,
submitted to the International Seminar ``Quarks-94'', Vladimir (1994);
available in pdf and ps format through SPIRES.  \hfill\break
\bigskip
\noindent
[28]  M. Gell-Mann, General Status: Summary and Outlook, in
{\it Proceedings
of the XVI International Conference on High Energy Physics};
conference held at
the University of
Chicago and the National
Accelerator Laboratory,  Sept. 1972; see
Vol. 4, p. 333 of the proceedings published by the National Accelerator
Laboratory, Batavia, IL. \hfill\break
\bigskip
\noindent
[29]  O. W. Greenberg, {\it Phys. Rev. Lett.} {\bf 13}, 598 (1964).
\hfill\break
\bigskip
\noindent
[30]  K. G. Wilson, The Origins of Lattice Gauge Theory, hep-lat/0412043.
\hfill\break
\bigskip
\noindent
[31] H. Fritzsch, A Time Dependence of QCD, hep-ph/0411391. \hfill\break
\bigskip
\noindent
[32]  H. Fritzsch and M. Gell-Mann, Current Algebra:  Quarks and What
Else?, in
{\it Proceedings
of the XVI International Conference on High Energy Physics};
conference held at
the University of
Chicago and the National
Accelerator Laboratory, Sept. 1972; see
Vol. 2, p. 135 of the proceedings published by the National Accelerator
Laboratory, Batavia, IL. The quoted remark is on p. 140.  \hfill\break

\vfill\eject
\bigskip
\bye